\documentstyle[12pt]{article}
\begin{document}
\newcommand{\bin}[2]{\left(\!\!\begin{array}{c}#1\\#2\end{array}\!\!\right)}
\newcommand{\tr}{\bigtriangleup}
\newcommand{\free}{ m^2 - \partial\,^2}
\newcommand{\lb}{\lambda}
\newcommand{\be}{\begin{equation}}
\newcommand{\ee}{\end{equation}}
\newcommand{\ba}{\begin{eqnarray}}
\newcommand{\ea}{\end{eqnarray}}
\newcommand{\fre}{ \mu^2 - \partial\,^2}

\Large

\begin{center}
{\large \bf A non-perturbative method of calculation of Green functions}\\
\vspace*{5 true mm}
{\bf V.E. Rochev}\\
{\it Institute for High Energy Physics, 142284 Protvino, Moscow region, Russia} 
\end{center}
{\small{ {\bf Abstract.} A new method for non-perturbative calculation of Green functions in
quantum mechanics and quantum field theory is proposed.
The method is based on an approximation of Schwinger-Dyson equation for
the generating functional by exactly soluble equation in 
functional derivatives. Equations of the leading approximation and
the first step are solved for  $\phi^4_d$-model. At $d=1$
(anharmonic oscillator) the ground state energy is calculated.
The renormalization program is performed for the field theory at $d=2,3$.
At $d=4$ the renormalization of the coupling  involves a trivialization
of the theory.}

\section{Introduction and General Consideration}

During many years the construction of non-perturbative approximate
solutions remains the urgent problem of quantum field theory.

In present work a method for the construction of such approximations
is proposed. The method is based on the approximation of the Schwinger-Dyson
equation for the generating functional by a simple equation in functional
derivatives, which can be solved exactly. This solution is a foundation
for the linear iterative scheme. Each step of the scheme consists 
in solving a closed system of integral equations.

The leading approximation and the first step are investigated
by the method for the $\phi^4_d$-model. At $d=1$, when the model 
corresponds to the anharmonic oscillator, a formula for the ground state energy
is obtained. At $d=2,3$ (super-renormalizable field theory) the renormalization
of the leading approximation and the first step is performed.
At $d=4$ (strictly renormalizable case) the renormalization of the coupling
leads to a non-physical singularity of the amplitude. This is a reflection
of the well-known triviality problem for the $\phi^4_4$-theory in a
non-perturbative region.

Consider the  theory of a scalar field $\phi(x)$ in Euclidean space
$(x \in E_d)$ with the action
\be
S(\phi) = \int dx \{\frac{1}{2} (\partial_{\mu}\phi)^2
 +\frac{m^2}{2} \phi^2 + \lb\phi^4\}.
\ee
The generating functional of  $ 2n$-point Green functions can be written as 
\be
G = \sum_{n=0}^{\infty} {G^{2n} \eta^n}, 
\ee
where $\eta(x,y)$ is a bilocal source. The $n$th derivative of G over
$\eta$ with the source being switched off is the $2n$-point Green function
$G^{2n}$.

Using of the bilocal source is an essential point of the
scheme presented, therefore we shall consider the theory without
spontaneous symmetry breaking with $m^2 > 0.$ A brief discussion
of the theory with a single source and with spontaneous symmetry
breaking is contained at the last section.
At $d \geq 2$   corresponding counterterms should be included
in the action for the cancellation of  ultraviolet divergences.

A simple way to obtain 
 Schwinger-Dyson equation for the generating functional  
with bilocal source $\eta(x,y)$ is
using the usual Schwinger-Dyson equation with single source $j(x)$
at  the presence of bilocal source $\eta$
\be
4\lb \frac{\delta^3 G}{\delta j^3(x)} + (\free) \frac{\delta G}{\delta j(x)} =
j(x) G + 2 \int  dx' \eta (x,x') \frac{\delta G}{\delta j(x')}.
\ee 
Differentiating of equation (3) over $j(y)$ and using the connection
condition
\ba
\frac{\delta^2 G}{\delta j(x) \delta j(y)} \nonumber
= \frac{\delta G}{\delta \eta (y,x)}\nonumber
\ea
we obtain after switching off the single source $j$ the following equation 
\be 
4\lb \frac{\delta^2 G}{\delta\eta(y,x)\delta\eta(x,x)} 
+ (\free)\frac{\delta G}{\delta\eta(y,x)}
- 2 \int  dx'\eta(x,x')\frac{\delta G}{\delta\eta(y,x')} - \delta(x-y) G = 0.
\ee
which contains the source $\eta$ only.

An idea of the presented iterative scheme is as follows: 
 we shall consider "an equation with constant coefficients"
as a leading approximation,
i.e.,  equation (4) with the next-to-last term omitted. This term contains
the source $\eta$ manifestly. The Green functions are the derivatives
of $G(\eta)$ in zero and only the behaviour of $G$ near $\eta = 0$ is essential,
therefore such an approximation seems to be acceptable.
 The equation of the leading  approximation will be
\be
4\lb \frac{\delta^2 G_0}{\delta\eta(y,x)\delta\eta(x,x)} 
+ (\free)\frac{\delta G_0}{\delta\eta(y,x)}
 - \delta(x-y) G_0 = 0.
\ee
The term omitted  contains the source and should be treated as a
perturbation.
Hence, the iteration procedure for the generating functional
\be
G = G_0 + G_1 + \cdots + G_n + \cdots
\ee
consists in the step-to-step solution of the equations
\be
4\lb \frac{\delta^2 G_n}{\delta\eta(y,x)\delta\eta(x,x)} 
+ (\free)\frac{\delta G_n}{\delta\eta(y,x)}
 - \delta(x-y) G_n =
2 \int dx'\eta(x,x')\frac{\delta G_{n-1}}{\delta\eta(y,x')}. 
\ee 
The solution of the leading approximation equation (5) is a functional
\be
G_0 = \exp \int dx dy \eta(y,x)\tr_0(x-y),
\ee
where $\tr_0$ is a solution of the equation
\be
4\lb\tr_0(0)\tr_0(x-y) + (\free)\tr_0(x-y) = \delta(x-y).
\ee
At $d\geq2$ the quantity $\tr_0(0)$ must be considered as
some regularization.

Equation (9) looks as the self-consistency equation, but differs
in the coefficient at $\lb$: in the self-consistency equation the 
coefficient is three times greater. In this sense  equation (9) is more
similar to the equation for the propagator in the leading approximation
of $1/N$-expansion. Certainly the similarity is completely superficial,
since the principle of the construction of the approximation scheme is
 different.

The solution of equation (9) is a free propagator
\be
\tr_0 = \frac{1}{\mu^2-\partial^2}
\ee
with the renormalized mass $\mu^2=m^2+4\lb\tr_0(0)$. The quantity $\tr_0(0)$
is defined from the self-consistency condition.

The propagator is a first derivative
of $G(\eta)$ over the source $\eta :\,\,\,
\tr={\frac{\delta G}{\delta\eta}}\mid_{\eta\, =0}$.
As can be easily  seen,
it is simply $\tr_0$  for the leading approximation. 

Notice, that all higher Green functions of the leading approximation
starting with four-point function
$G^4={\frac{\delta^2 G}{\delta\eta^2}}\mid_{\eta\, =0}$ 
do not possess the correct connected structure and, 
correspondingly, the complete bose-symmetry. The correct connected
structure and other consequences  of bose-symmetry (e.g., crossing etc.)
will be  restored in consecutive order at  following steps of
the iteration scheme. This is easy to see, for example, 
 analysing the iteration
scheme equations at $\lb\rightarrow 0$. Such a peculiarity of the
iteration scheme is originated by bilocal source and is not something
exceptional : as is well-known, the similar phenomenon appears also
in constructing $1/N$-expansion in the bilocal source formalism.

The first step equation for the generating functional $G_1$ reads
\be
4\lb \frac{\delta^2 G_1}{\delta\eta(y,x)\delta\eta(x,x)} 
+ (\free)\frac{\delta G_1}{\delta\eta(y,x)}
 - \delta(x-y) G_1 =
2 \int dx' \eta(x,x')\tr_0(x'-y) G_0 
\ee
A solution of equation (11) is looked for in the form $G_1=P_1(\eta)G_0$,
where\\ $P_1=\frac{1}{2}F\eta^2+\tr_1\eta$.
Taking into account the leading approximation, equation (11)
gives a system of equations for $F$ and $\tr_1$:
\ba
(\fre_x)F\!\bin{x\,y}{x'\,y'} + 4\lb F\!\bin{x\,x}{x'\,y'}\tr_0(x-y)
= \nonumber\\
=\delta(x-y')\tr_0(x'-y)+\delta(x-x')\tr_0(y-y'),
\ea
\be
(\fre)\tr_1(x-y)+4\lb\tr_1(0)\tr_0(x-y)+4\lb F\!\bin{x\,x}{x\,y} = 0.
\ee
Equation (12) is the linear integral equation for the function $F$
in the momentum space.
A solution of the equation is
\ba
F\!\bin{x\,y}{x'\,y'} = 
\tr_0(x-y')\tr_0(x'-y)+\tr_0(x-x')\tr_0(y-y')\nonumber\\
-4\lb \int dx_1 dy_1 \tr_0(x-x_1)\tr_0(y-x_1)K(x_1-y_1)\tr_0(y_1-x')\tr_0(y_1-y'),
\ea
where the kernel $K$ is a solution of the equation
\be
K(x-y) = 2\delta(x-y) - 4\lb \int dx' L(x-x')K(x'-y),
\ee
and $L(x-y) \equiv \tr^2_0(x-y)$ is a single loop. Equation (15)
 can be easily solved in the momentum space. 
 Its solution  is
\be
\tilde K (p) = \frac{2}{1+ 4\lb \tilde L (p)}.
\ee
Notice, that first two terms in formula (14) for $F$ are the missed
connected structure of the four-point function of leading approximation.
Hence the connected structure of four-point function is restored
yet at the first step of iterations. 

To solve equation (13) for $\tr_1$ is also quite simple.
Taking into account the formulae above the solution can be written as
\be
\tr_1(x-y) = - \int dx' dy' \tr_0(x-x')\Sigma_r(x'-y')\tr_0(y'-y),
\ee
where
\be
\Sigma_r(x-y) = [4\lb\tr_1(0)+8\lb\tr_0(0)]\delta(x-y)+\Sigma(x-y),
\ee
\be
\Sigma(x-y) = -(4\lb)^2\tr_0(x-y) \int dx' L(x-x') K(x'-y).
\ee
The quantity $\tr_1(0)$ is defined by the self-consistency condition.

At $\lb \rightarrow 0$, as is easy to see,
$\tr = \tr_0 + \tr_1 = \tr^{pert} +  O(\lb^2)$,
where $\tr^{pert}$ is the propagator of the perturbation theory, i.e. at small
$\lb$ the first step propagator reproduces correctly the first term
of the usual perturbation theory in the coupling.

In the general case, the solution of  equation (7) for the $n$-th step
 of the iteration scheme is
\be
G_n = P_n(\eta) G_0,
\ee
where $P_n$ is a polynomial in $\eta$ of a degree $2n$.
Therefore at the $n$-th step the computation of Green functions reduces to
 solving  a system of $2n$ linear integral equations.

\section{Anharmonic Oscillator}

At  $d=1$ the model  with the action (1) describes the quantum-mechanical
anharmonic oscillator. The parameter $m^2$ corresponds in the case to  a
frequency of a harmonic oscillator described by quadratic terms. At $d=1$
ultraviolet divergences are absent, and quantities $\tr_0(0)$
and $\tr_1(0)$ are finite. Consequently, formulae (8) - (19) are applied
directly for the computation of Green functions.

Since at $d=1\,\,\,\,\,\tr_0(0) = \frac{1}{2\mu}$, the self-consistency
condition becomes the equation for a renormalized mass (or, more exactly,
for a "renormalized frequency")   $\mu^2$:
\be
\mu^2 = m^2 + \frac{2\lb}{ \mu}.
\ee
Here $\mu=\sqrt{\mu^2}$.
The equation  has the unique positive  
solution at all positive $m^2$ and $\lb$.

To calculate a ground state energy $E$ one can use the well-known
formula (see, for example, \cite{1,2})
\be
\frac{dE}{d\lb} = G^4 (0,0,0,0).
\ee
In this formula $G^4$ is the four-point (or, two-particle) function:
$G^4=\frac{1}{G}{\frac{\delta^2 G}{\delta\eta^2}}\mid_{\eta\, =0}$. 
With the formulae above for Green functions of the leading approximation
and the first step, we obtain the following formula for the ground
state energy of the anharmonic oscillator:
\be
\frac{dE}{d\lb} = \frac{1}{4\mu^2} + \frac{1}{\mu M}(1- 
\frac{2\lb}{\lb + \mu^3}(1- \frac{2\lb}{\mu(M+2\mu)^2})).
\ee 
Here $ M=\sqrt{4\mu^2+\frac{4\lb}{\mu}}$. Integrating the formula
with  a boundary condition $E\mid_{\lb=0}~=~m/2$ taken into account, 
one can calculate the ground state energy for all values of the coupling:
$0\,\leq\,\lb\,<\,\infty$. 

At  $\lb\rightarrow 0:\,\,\,
E=m(\frac{1}{2} + \frac{3}{4}\frac{\lb}{m^3} + O(\lb^2))$
--- the perturbation theory is reproduced up to second order.

At $\lb\rightarrow\infty:\,\,\,E=\epsilon_0\lb^{1/3} + O(\lb^{-1/3})$,
and $\epsilon_0 = 0.756$. The coefficient $\epsilon_0$ differs by 13\% from
the exact numerical one $\epsilon_0^{exact} = 0.668$ 
(see, for example, \cite{3}).

At $\lb/m^3 = 0.1$ the result of the calculation with formula (23)
differs from the exact numerical one  \cite{3} by 0.8\%, and at
$\lb/m^3 = 1$ differs by 6.3\%.

Consequently, the first step formula (23) approximates the ground
state energy for all values of $\lb$ with the accuracy that varies
smoothly from 0 (at $\lb\rightarrow 0$) to 13\% (at $\lb\rightarrow\infty$).
Comparing these results with the results of other approximate methods,
in particularly, with the method of variational perturbation theory (VPT)
\cite{1,2} or the  $\delta$-expansion method \cite{4}, we can see that
this method gives best results for the intermediated coupling region
$\lambda \sim 0.1 m^3$. Really, at $\lambda/m^3 = 0.1$ the accuracy
of first step of the calculations is better than the
accuracy of fifth step of VPT (see Table 6 of ref. \cite{2})
and five times as higher in comparison with the usual perturbation
theory in the order ${\cal O}(\lambda^5)$. However, for the asymptotic
region of strong coupling the accuracy of the given calculations
is not so good, and other methods such as VPT give more exact results
at $\lambda\rightarrow\infty$. It seems be likely that a combination of the 
method proposed with the methods of VPT type would gives good result for
all coupling values.

\section{Super-renormalizable theory (d=2 and d=3)}

At $d\geq 2$ the action (1) should be added by  counterterms for 
the elimination of ultraviolet divergences. First consider super-renormalizable
theory ($d=2$ and $d=3$). It is sufficient to add counterterms of 
mass renormalization  $\frac{\delta m^2}{2}\phi^2$ and wave function
renormalization $\frac{\delta z}{2}(\partial_{\mu} \phi)^2$ in the case. 
The Schwinger-Dyson equation has the form of equation (4) with the substitution
\be
m^2\rightarrow m^2 + \delta m^2,\,\,\,\,\partial^2\rightarrow (1+\delta z)
\partial^2
\ee 
There is no need to add a counterterm of wave function renormalization
for the leading approximation, and the equation of the leading
approximation will be
\be
4\lb \frac{\delta^2 G_0}{\delta\eta(y,x)\delta\eta(x,x)} 
+ (\delta m^2_0 + \free)\frac{\delta G_0}{\delta\eta(y,x)}
 - \delta(x-y) G_0 = 0.
\ee
The l.h.s. of iteration scheme equation (7) is of the same form as the l.h.s.
of equation (25) is.
At $n\geq 1$ the counterterms $\delta m^2_n$ and $\delta z_n$ should be
considered as perturbations. Therefore corresponding terms should be
added to the r.h.s. of  equation (7). So, the first step equation
will be
\ba
4\lb \frac{\delta^2 G_1}{\delta\eta(y,x)\delta\eta(x,x)} 
+ (\delta m^2_0 + \free)\frac{\delta G_1}{\delta\eta(y,x)}
 - \delta(x-y) G_1 =\nonumber\\
= 2 \int dx' \eta(x,x')\frac{\delta G_{0}}{\delta\eta(y,x')} - 
\delta m^2_1\frac{\delta G_{0}}{\delta\eta(y,x)}
+ \delta z_1 \partial^2\frac{\delta G_{0}}{\delta\eta(y,x)}.
\ea
  
The normalization condition on the physical renormalized mass $\mu^2$ 
gives us  a counterterm of  the mass renormalization in the leading
approximation
\be
\delta m^2_0 = \mu^2 - m^2 - 4\lb\tr_0(0).
\ee
This counterterm diverges logarithmically  at $d=2$ and linearly at $d=3$.

For the first step of the iteration  equation (12) for $F$ remains unchanged,
and its solution is described by the same formulae (14)-(16). The equation
for $\tr_1$ changes in correspondence with equation (26). Its solution
can be written in the same form (17), but now for $\Sigma_r$ one gets
\be
\Sigma_r = 4\lb \tr_1(0) + \delta m^2_1 - 2\delta m^2_0 
- \delta z_1 \partial^2 + \Sigma,
\ee
where $\Sigma$ is given by formula (19). The normalization conditions
\be
\tilde \Sigma_r(-\mu^2) = 0,\,\,\,\,\,\tilde \Sigma_r'(-\mu^2) = 0
\ee
give us the counterterms of the first step
\ba
\delta z_1 = - \tilde \Sigma'(-\mu^2),\nonumber\\
\delta m^2_1 = 2\delta m^2_0 - 4\lb\tr_1(0) - \tilde \Sigma (-\mu^2)
- \mu^2\tilde \Sigma'(-\mu^2).
\ea
The renormalized mass operator is
\be
\tilde \Sigma_r (p^2) = \tilde \Sigma (p^2) - \tilde \Sigma (-\mu^2)
- (p^2 +  \mu^2)\tilde \Sigma'(-\mu^2),
\ee
where, in correspondence with (19),
\be
\tilde \Sigma (p^2) = -(4\lb)^2 \int \frac{dq}{(2\pi)^d}
\frac{1}{\mu^2 + (p-q)^2} \frac{2 \tilde L(q^2)}{1+ 4\lb \tilde L(q^2)}.
\ee
The counterterm $\delta z_1$ is finite at $d=2,3$. The counterterm 
$\delta m^2_1$ diverges 
as that  of the leading approximation does,
 namely, logarithmically at $d=2$ and linearly at $d=3$.
As the simple loop $\tilde L(p^2)$ behaves for $p^2 \rightarrow \infty$ as 
$\frac{1}{p^2} \log \frac{p^2}{\mu^2}$ at  $d=2$ and  
$\frac{1}{\sqrt{p^2}}$ at $d=3$,
 the integral (32) for $ \tilde \Sigma (p^2)$ converges
at $d=2$ and diverges logarithmically at $d=3$. Surely the renormalized
mass operator (31) is finite in any case.

A part of  "redundant" subtractions  in formula (31) completely unclear
from the point of view of the usual perturbation theory divergences.
 This part clears at 
the strong coupling limit  in formulae (31)-(32).
Really, at $\lb\rightarrow\infty$
\be
\tilde \Sigma (p^2) = const +  \int \frac{dq}{(2\pi)^d}
\frac{1}{\mu^2 + (p-q)^2} \frac{2}{ \tilde L(q^2)} + O(\frac{1}{\lb}).
\ee
The integral in equation (33) diverges {\it quadratically} at $d=2,3$.
Hence, it becomes clear that the "redundant" subtractions conserve the
finiteness of renormalized theory in the strong coupling limit.

\section{Strictly renormalizable theory (d=4)}
 
At  $d=4$ besides the renormalizations of the mass and the wave function a
coupling  renormalization is necessary. Therefore simultaneously
with the substitution (24)  the substitution
 $\lb\rightarrow\lb+\delta\lb$ is also needed
in the Schwinger-Dyson equation (4). The leading approximation
equation will be
\be
4(\lb+\delta\lb_0) \frac{\delta^2 G_0}{\delta\eta(y,x)\delta\eta(x,x)} 
+ (\delta m^2_0 + \free)\frac{\delta G_0}{\delta\eta(y,x)}
 - \delta(x-y) G_0 = 0.
\ee
Due to the presence of the counterterm $\delta\lb$ the normalization
condition on the renormalized mass $\mu^2$ for the leading approximation
becomes a connection between counterterms $\delta m^2_0$ and $\delta\lb_0$
\be
\delta m^2_0 + 4(\lb + \delta\lb_0)\tr_0(0) = \mu^2 - m^2.
\ee
As we shall
see below, counterterm $\delta\lb_0$ (and, consequently,  
 $\delta m^2_0$) will be fixed on the {\it following} step
of the iteration scheme.

The first step equation will be of the form  (26)
with the substitution $\lb\rightarrow\lb+\delta\lb_0$ 
in the l.h.s. and with an additional term 
$-4\delta\lb_1\cdot\delta^2 G_0 / \delta\eta(y,x)\delta\eta(x,x)$
in the r.h.s. The equation for $F$ will  differ
from  equation (12) by the substitution $\lb\rightarrow\lb+\delta\lb_0$ only.
Therefore the formulae for its solution also will differ
from  formulae (14)-(16) by the same substitution.
At $d=4$ the single-loop integral $\tilde L(p^2)$ diverges logarithmically,
and the renormalization of the coupling is necessary. Let define a
two-particle amplitude --- the amputated connected part of the
four-point function
\be
A = \tr_0^{-1}\tr_0^{-1}F^{con}\tr_0^{-1}\tr_0^{-1}.
\ee
Here a multiplication by $\tr_0^{-1}$ is understood in the operator sense.
$F^{con}$ is the connected part of $F$. It is easy to see that the amplitude
depends from a variable $p=p_x+p_y$ only and has the form
\be
\tilde A(p^2) = -\frac{8(\lb+\delta\lb_0)}{1+4(\lb+\delta\lb_0)\tilde L(p^2)}.
\ee
Define a renormalized coupling $\lb_r$ as a value of the amplitude in
a normalization point
\be
\tilde A(M^2) = -8\lb_r =
 -\frac{8(\lb+\delta\lb_0)}{1+4(\lb+\delta\lb_0)\tilde L(M^2)}.
\ee
From  equation (38) one obtains a counterterm of  coupling
renormalization
\be
\delta\lb_0 = - \lb + \frac{\lb_r}{1-4\lb_r \tilde L(M^2)}
\ee
and renormalized amplitude
\be
\tilde A(p^2) = -\frac{8\lb_r}{1+4\lb_r\tilde L_r(p^2; M^2)},
\ee
where $\tilde L_r(p^2; M^2) = \tilde L(p^2) - \tilde L(M^2)$ is
a renormalized loop that possesses a finite limit at the regularization removing.

Taking the renormalization of the two-particle amplitude in  such a manner,
one can  solve the equation for $\tr_1$ and  renormalize the mass operator
in correspondence with the general principle of normalization on the physical
mass (see equation (29)).
 But in four-dimensional
case one gets an essential obstacle. At the regularization removing, 
$\delta\lb_0\rightarrow -\lb.$ It is evident from  equation (39).
Therefore the coefficient $\lb+\delta\lb_0$ in the leading approximation
equation (34) vanishes. The same is true for  all the 
subsequent iterations.
The theory is trivialized. One can object that an expression
\be 
(\lb+\delta\lb_0)\cdot \frac{\delta^2 G}{\delta\eta(y,x)\delta\eta(x,x)}
\ee
is really an indefinite quantity  of  $0\cdot\infty$ type,
 and the renormalization is,
in the essence, a definition of the quantity. But it does 
not save a situation in
this case since the renormalized amplitude (40) possesses a non-physical
singularity in a deep-euclidean region (it is a well-known Landau pole).
The unique noncontradictory possibility is a choice $\lb_r\rightarrow 0$
at the regularization removed. This is the trivialization of the theory again.
This trivialization appears almost inevitably in an investigation
of $\phi^4_4$-theory beyond  the perturbation theory and is a practically
rigorous result (see \cite{5}). 
Notice, that  contrary to perturbation theory which is absolutely
nonsensitive to the triviality of the theory, the method proposed leads
to the trivialization already at the first step.

\section{Single Source}

The method considered above
is based essentially on the bilocality of the source. Since the 
bilocal source produces  $2n$-point functions only, the method
cannot be applied in its present form to a theory with 
spontaneous symmetry breaking when 
\be
<0\mid \phi \mid 0>\neq 0. 
\ee
For description of the spontaneous symmetry breaking it is necessary
to switch on a single source $j(x)$, i.e. to consider
Schwinger-Dyson  equation (3). 

Consider a theory with the single source $j$ and with
the bilocal source switched off: $\eta = 0$. Schwinger-Dyson
equation for the generating functional $G(j)$ is
\be
4\lb \frac{\delta^3 G}{\delta j^3(x)} + (\free) \frac{\delta G}{\delta j(x)} =
j(x) G.
\ee

Let apply to  equation (43) the same idea of approximation 
 by the equation with "constant" coefficients, i.e. consider
as a leading approximation the equation
\be
4\lb \frac{\delta^3 G_0}{\delta j^3(x)}
 + (\free) \frac{\delta G_0}{\delta j(x)} = 0.
\ee 
Then an iteration scheme will be described by the equation
\be
4\lb \frac{\delta^3 G_n}{\delta j^3(x)}
 + (\free) \frac{\delta G_n}{\delta j(x)} =
j(x) G_{n-1}.
\ee

The leading approximation equation (44) has a solution
\be
G_0 = \exp \{ \int dx v(x)j(x)\}.
\ee

Surely $v$ does not depend on $x$ in a translational invariant theory.
Therefore an equation for $v$ will be
\be
4\lb v^3 + m^2 v = 0.
\ee

At $m^2 \geq 0,\,\,\lb>0$ the equation has unique real-valued solution
 $v=0$, that corresponds to leading approximation $G_0 = 1$.
Iteration scheme (45) based on this leading approximation coincides
with the perturbation theory in the coupling --- the leading approximation
is too simple and does not contain any nonperturbative effects.

At $m^2 < 0$ besides of this solution the following real-valued ones exist
\be
v = \pm \sqrt{-\frac{m^2}{4\lb}},
\ee
which correspond to spontaneous breaking of a discrete symmetry ($P$-parity) of
$\phi^4$-theory. A calculation of the ground state energy 
based on  the formulae of equation (22)
type demonstrates that the state with spontaneous symmetry breaking
is energetically preferable and so it is a physical vacuum of the theory
at $m^2 < 0$. There is the way to describe the leading non-perturbative
effect, i.e. spontaneous symmetry breaking, by the method. 

A first step equation with counterterms will be
\ba
4\lb \frac{\delta^3 G_1}{\delta j^3(x)}
 + (\free) \frac{\delta G_1}{\delta j(x)} =\nonumber\\
=j(x) G_{0} - 4\delta\lb_1 \frac{\delta^3 G_0}{\delta j^3(x)}
- \delta m^2_1 \frac{\delta G_0}{\delta j(x)}
+ \delta z_1 \partial^2 \frac{\delta G_0}{\delta j(x)}=\nonumber\\
= (j(x) - 4\delta\lb_1 v^3 - v \delta m^2_1) G_0.
\ea
A solution of  equation (49) should be looked for as
$G_1 = P_1(j) G_0$, where\\ $P_1(j) = \frac{1}{2} \tr_1 j^2 + \Phi_1 j$.
The equation for $\tr_1$ with taking into account formulae of leading
approximation (46)-(48) has a solution
\be
\tr_1 = \frac{1}{\fre},
\ee
where $\mu^2 = -2m^2 > 0$. The reconstruction of the vacuum leads 
to the corresponding
reconstruction of the one-particle spectrum. All the picture exactly
corresponds to a description in the effective potential language, but
the conception of  the effective potential is not  used  at all.
Subsequent iterations leads to the renormalized perturbation theory
over the physical nonsymmetrical vacuum. A remarkable feature of the scheme
is an absence of symmetry breaking counterterms even at intermediate steps
of calculations. For the ultraviolet divergences removing the
counterterms $\delta m^2, \delta\lb$ and $\delta z$
are sufficient.

In conclusion notice, though at $m^2 > 0$ values of $v$ in equation (48) are
imagine, the corresponding real-valued  solutions of equation (44) exist,
for example
\be
G_0 = \cos \{ w\int dx j(x) \},
\ee
where $w^2 = m^2/4\lb$. At first step of the iteration such a leading
approximation gives tachyons and so it is physically unacceptable.
It is possible that similar solutions can be useful for an investigation
of the problem of spontaneous symmetry breaking in $\phi^4_2$-theory
with $m^2 > 0$ (see, for example \cite{6}).
 Of course,   the computational scheme should be modified in that case.
 
\section*{Aknowlegements} 
Author is grateful to A.I. Alekseev, B.A. Arbuzov and I.L. Solovtsov
for useful discussion. The work is supported by Russian Foundation
for Basic Researches, grant No.95-02-03704.

\end{document}